# A QoE-Based Scheduling Algorithm for UGS Service Class in WiMAX Network

Tarik Anouari, Abdelkrim Haqiq

*Abstract*— To satisfy the increasing demand for multimedia services in broadband Internet networks, the WiMAX (Worldwide Interoperability for Microwave Acces) technology has emerged as an alternative to the wired broadband access solutions. It provides an Internet connection to broadband coverage area of several kilometers in radius by ensuring a satisfactory quality of service (QoS), it's an adequate response to some rural or inaccessible areas. Unlike DSL (Digital Subscriber Line) or other wired technology, WiMAX uses radio waves and can provide point-to-multipoint (PMP) and point-to-point (P2P) modes. In parallel, it's observed that in the opposite of the traditional quality evaluation approaches, nowadays, current researches focus on the user perceived quality, the existing scheduling algorithms take into account the QoS and many other parameters, but not the Quality of Experience (QoE). In this paper, we present a QoE-based scheduling solution in WiMAX network in order to make the scheduling of the UGS connections based on the use of QoE metrics. Indeed, the proposed solution allows controlling the packet transmission rate so as to match with the minimum subjective rate requirements of each user. Simulation results show that by applying various levels of mean opinion score (MOS) the QoE provided to the users is improved in term of throughput, jitter, packet loss rate and delay.

## I. INTRODUCTION

Usually, the network was examined objectively by evaluating a number of parameters to evaluate the quality of network service. This evaluation is known as the QoS of the network, it refers to the ability of the network to obtain a more deterministic performance, and therefore data can be transported with a minimum packet loss, minimum delay and maximum throughput. The QoS does not take into account the user's perception of the service provided. Another approach which takes into account the user's perception is known as QoE, it's a subjective evaluation that associates human dimensions; it groups together user perception, expectations, and experience of application and network performance.

In order to understand the quality as perceived by end users, QoE has become a very active area of research. Many related works were published on analyzing and enhancing QoE [12] in WiMAX network. The study in [14] proposed an estimation method of QoE metrics based on QoS metrics in WiMAX network. The QoE was evaluated by using a Multilayer Artificial Neural Network (ANN).The results show an efficient estimation of QoE metrics with respect to QoS parameters.

Manuscript received on March, 2014.
**Tarik Anouari**, Computer, Networks, Mobility and Modeling laboratory/ Department of Mathematics and Computer/ FST, Hassan 1st University, Settat, Morocco/ E-NGN Research group, Africa and Middle East.
**Abdelkrim Haqiq**, Computer, Networks, Mobility and Modeling laboratory/ Department of Mathematics and Computer/ FST, Hassan 1st University, Settat, Morocco/ E-NGN Research group, Africa and Middle East.

Other works like [6, 7 and 8] also focus on the ANN method to adjust the input network parameters to get the ideal output to satisfy end users. Principally, the success of the ANN approach depends on the model's capacity to completely learn the nonlinear interactions between QoE and QoS. In [16], Muntean presents a learner QoE model that in addition to the user-related content adaptation, considers delivery performance-based content personalization in order to improve user experience when interacting with an online learning system. Simulation results demonstrate significant improvements in terms of learning achievement, learning performance, learner navigation and user QoE

In [3], our study was focused on analyzing QoS performances of VoIP and Video traffic using different service classes with respect to QoS parameters such as throughput, jitter and delay. The simulation results show that UGS service class is the best suited to handle VoIP traffic. This paper proposes a novel approach based on the user perception of Quality to provide best WiMAX network performances especially for the real-time traffic. The target of this improvement is to schedule traffic of UGS service class.

The rest of this paper is organized as follows. Section 2 gives a short description of the WiMAX technology. In section 3, a QoE overview background is presented. The proposed QoE-based scheduling algorithm is described in detail in the section 4. Simulation environment and performance parameters are described in Section 5. Section 6 shows simulation results and analysis. Finally, section 7 concludes the paper with future work direction.

## II. WIMAX TECHNOLOGY

WiMAX is a standard wireless metropolitan area network created by the Intel and Alvarion companies in 2002 and ratified by the IEEE (Institute of Electrical and Electronics Engineer) as IEEE-802.16 [10, 11]. More precisely, WiMAX is the commercial label delivered by the WiMAX Forum for equipment compliant with the IEEE 802.16 standard to ensure a high level of interoperability between different equipment. The objective of WiMAX is to provide a broadband internet connection on a coverage area of several kilometers. Thus, in theory, WiMAX provides data rates of 70 Mbit/s with a range of 50 kilometers.

WiMAX can be used in PMP mode, in which from a central base station, serving multiple client terminals is ensured and in P2P mode, in which there is a direct link between the central base station and the subscriber. PMP mode is less expensive to implement and operate while P2P mode can provide greater bandwidth.

### A. QoS in WiMAX Network

The QoS was introduced natively in WiMAX [18]. It may satisfy QoS requirements for a wide range of services and data applications especially with the high speed connection, asymmetric capabilities UL and DL, flexible mechanisms for resource allocation. Some services are very demanding, VoIP cannot tolerate delay in the transmission of data.

The concept of QoS clearly depends on the service considered, its requirement of response time, which is its sensitivity to transmission errors... etc. For video streaming, we will need a near real-time transfer, with very low latency and low jitter, while VoIP traffic is intolerant of network delays and retransmission.

Respecting QoS requirement becomes very important in IEEE802.16 systems to guarantee their performance, in particular in the presence of various kinds of connections, namely the current calls, new calls and the handoff connection.

### B. WiMAX Network Architecture

WiMAX runs in infrastructure mode, it typically consists of a base station named BTS (Base Transceiver Station) or BS (Base Station), which is the central control element of the network and at least one station (SS -Subscriber Station). The BS can provide various levels of QoS over its queuing, scheduling, control, signaling mechanisms, classification and routing. Figure 1 shows the architecture of WiMAX network [10, 11].

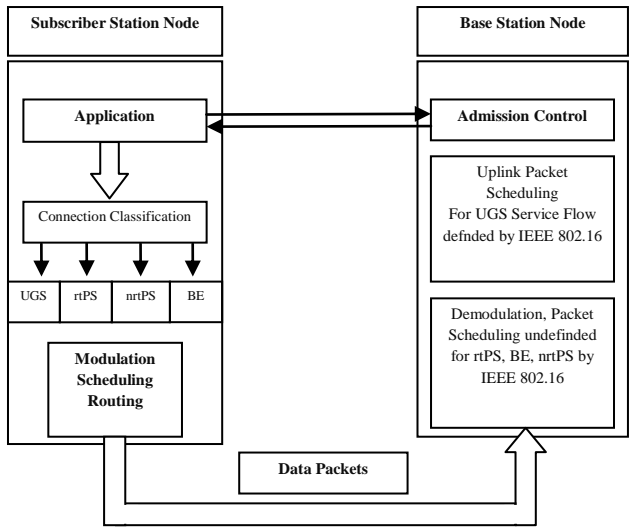

Figure 1: WiMAX Network Architecture

### C. Different Service Classes in WiMAX

Several types of traffic may be considered. QoS is negotiated at the service flow during the connection establishment. Each connection on the uplink (Uplink) is mapped to a specific service. Each service is associated with a set of rules imposed by the scheduler of the BS responsible for assigning the capacity of the uplink and other parameters between SS and BS. The services provided by WiMAX are classified according to the parameters defining the quality of service associated with a connection. Four classes of service are available in the IEEE 802.16-2004 standard [10], Best Effort (BE), Real-Time Polling Service (rtPS), Non-Real Time Polling Service (nrtPS) and Unsolicited Grant Service (UGS). The ertPS service class was added specifically for the mobile version [1].

Some services are very demanding in QoS, while others have fewer requirements. Table 1 classifies different service classes of WiMAX and gives their description and QoS parameters.

Table 1: Service classes in WiMAX

| Service | Description | QoS parameters |
|---|---|---|
| UGS | Real-time data streams comprising fixed size data packets at periodic intervals | Maximum Sustained Rate<br>Maximum Latency Tolerance<br>Jitter Tolerance |
| rtPS | support real-time service flows that periodically generate variable-size data packets | Traffic priority<br>Maximum latency tolerance<br>Maximum reserved rate |
| ertPS | Real-time service flows that generate variable-sized data packets on a periodic basis. | Minimum Reserved Rate<br>Maximum Sustained Rate<br>Maximum Latency Tolerance<br>Jitter Tolerance<br>Traffic Priority |
| nrtPS | Support for non-real-time services that require variable size data grants on a regular basis | Traffic priority<br>Maximum reserved rate<br>Maximum sustained rate |
| BE | Data streams for which no data minimum service level is required. | Maximum Sustained Rate<br>Traffic Priority |

### III. QUALITY OF EXPERIENCE

Quality of Experience (QoE, user Quality of Experience or simply QX) is the measure perceived by the user on the service provided. The idea of QoE monitoring solutions is relatively innovative as it focuses on the perception of the end user to ensure that he is satisfied.

### A. Quality of Experience vs Quality of Service assessment

Recent years have seen a huge technological advancement in the field of packet networks. The internet as a part of this class of systems, have seen the birth of many multimedia applications. Various services such as Internet TV, video on demand, internet radio, multimedia data, IP telephony or teleconferencing have become our daily attraction and represent a large sector of the telecommunications market and an active area of research. Since that time the acronym QoS has been used to describe the improved performance realized by hardware and / or software. But with the rapid evolution of streaming video and VoIP from one year to another, the metrics of the QoS such as bandwidth, delay, jitter and packet loss which are generally used to ensure the services fail to measure subjectivity associated with human perception and thus was born the QoE, which is a measure of personal judgment of the user according to his experience. Indeed, the notion of user experience has been introduced for the first time by Dr. Donald Norman, citing the importance of designing a user [17] centered service.

Gulliver and Ghinea [9] classify QoE into three components: assimilation, judgment and satisfaction. The assimilation is a quality measure of the clarity of the contents by an informative point of view. The judgment of quality reflects the quality of presentation. Satisfaction indicates the degree of overall assessment of the user.

QoE and QoS are two complementary concepts: QoE indicators are used to monitor the quality offered to users and QoS indicators to identify and analyze the causes of network congestion. These two solutions used in parallel are a complete system monitoring.

*B. QoE Measurement approaches*

There are two main quality assessment approaches, namely objective and subjective performance evaluation. Subjective evaluation is carried out by end users who are asked to measure the perceived quality in a controlled environment. The most often used measurement is the MOS recommended by the International Telecommunication Union (ITU) [13], and it's defined as a numeric value evaluation from 1 to 5 (i.e. poor to excellent).

Objective approaches are usually based on algorithms, mathematical and/or comparative techniques that generate a quantitative measure of the service provided.

Peter and Bjørn [5] classify the existing approaches of measuring network service quality from a user perception into three classifications, namely: Testing User-perceived QoS (TUQ), Surveying Subjective QoE (SSQ) and Modeling Media Quality (MMQ). The first two approaches collect subjective information from users, whereas the third approach is based on objective technical assessment. Figure 2 [2] gives an overview of the classification of the existing approaches.

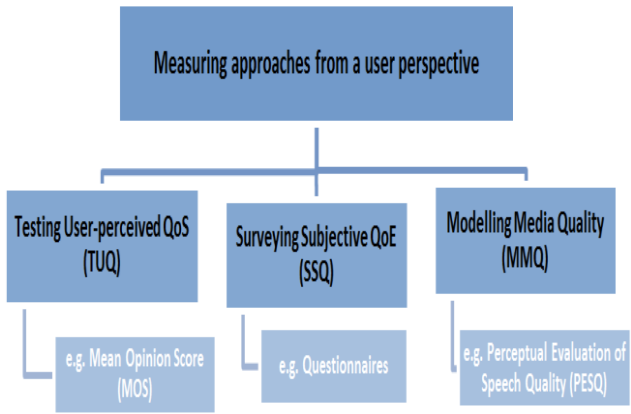

Figure 2: The approaches for measuring network service quality from a user perception

## IV. QOE-BASED SCHEDULING ALGORITHM MODEL

In this section we present a QoE-based scheduling algorithm to provide QoE to WiMAX network, since it's observed that the available scheduling algorithms take into account QoS but do not provide QoE, where virtually every user has different subjective requirement of the system.

*A. Proposed QoE-based scheduling algorithm model*

The proposed QoE-based scheduling algorithm is based on two QoE requirements, each user has an initial maximum transmission rate and a minimum subjective rate requirement. The scheduler works as follows, each node starts sending traffic with a maximum rate. When a packet loss occurs with a given user then the system check on each user if the transmission rate is higher than the minimum subjective requirement, in this case the transmission rate is reduced, otherwise the transmission continues at the same rate. The rate returns to the original maximum value during the simulation, it's rested every 20 seconds, we observe that it takes 18 seconds to all users to reach the minimal transmission rate.

Figure 3 shows the activity diagram of the proposed scheduling algorithm

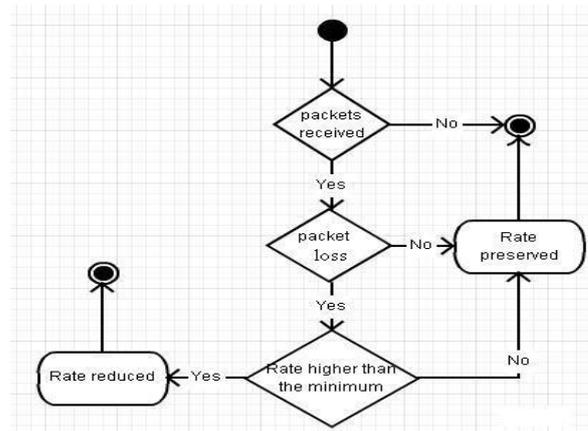

Figure 3: Activity diagram of the proposed QoE-based scheduling algorithm

## V. SIMULATION ENVIRONNEMENT

*A. Simulation Model*

In this paper, we analyze the performances of the proposed QoE-based scheduling algorithm, as we consider the Wireless-OFDM PHY layer, our QoE-based scheduling algorithm is compared with the famous WiMAX module developed by NIST (National Institute for Standards and Technologies), which is based on the IEEE 802.16 standard (802.16-2004) and the mobility extension (80216e-2005) [19], it provides a number of features including OFDM PHY layer. The Network Simulator (NS-2) [15] is used. Our simulation scenario consists of creating five wireless nodes (SS, subscriber stations) and connecting them to a BS. A sink node is created and attached to the base station to accept incoming packets. A traffic agent is created and then attached to the source node.

Finally, we set the traffic that produces each node. The first node has run with CBR (Constant Bit Rate) packet size of 200 bytes and interval of "0,0015", the second node has run with CBR packet size of 200 bytes and interval of "0,001", the third node has run with CBR packet size of 200 bytes and interval of "0,001", the fourth node has run with CBR packets size of 200 bytes and interval of "0,001" and fifth node has run with CBR packet size of 200 bytes and interval of "0,0015". The initial transmission rate that produces each node is about "133,3 Kbps", "200 Kbps", "200 Kbps", "200 Kbps" and "133,3 Kbps" respectively. All nodes have the same priority.

Each user has a minimum requirement, so the first user requires minimal traffic rate of "120 Kbps", the second "150 Kbps", the third "150 Kbps", the fourth "150 Kbps" and the fifth "120 Kbps".

The following table summarizes the above description about the produced and required traffic rate of each user.

Table 2: User's traffic parameters

| Users \ Traffic rate | Initial traffic rate (Kbps) | User minimum requirement (Kbps) |
|---|---|---|
| User 1 | 133,33 (200byte/0. 0015) | 120 |
| User 2 | 200 (200byte/0. 001) | 150 |
| User 3 | 200 (200byte/0. 001) | 150 |
| User 4 | 200 (200byte/0. 001) | 150 |
| User 5 | 133.33 (200byte/0. 0015) | 120 |

**A QoE-Based Scheduling Algorithm for UGS Service Class in WiMAX Network**

To perform this simulation, the network simulator NS-2.29 was used, we have implemented the QoS-included WiMAX module [4] within NS-2.29. This module is based on the NIST implementation of WiMAX [19], it consists of the addition of the QoS classes as well as the management of the QoS requirements.

The resulted trace files are interpreted and filtered based on a PERL script, it's an interpretation script software used to extract datas from trace files in term of throughput, packet loss rate, jitter and delay. The extracted analysis results are plotted in graphs using EXCEL software.

### B. Simulation Parameters

The same simulation parameters are used for both NIST and QOE-based scheduling algorithms, table 3 summarizes the simulation parameters:

Table 3: Simulation parameters

| Parameter | Values |
|---|---|
| Network interface type | Phy/WirelessPhy/OFDM |
| Propagation model | Propagation/OFDM |
| MAC type | Mac/802_16/BS |
| Antenna model | Antenna/OmniAntenna |
| Service class | UGS |
| Packet size | 200 bytes |
| Frequency bandwidth | 5 MHz |
| Receive Power Threshold | 2,025e-12 |
| Carrier Sense Power Threshold | 0,9 * Receive Power Threshold |
| Channel | 3,486e+9 |
| Simulation time | 200s |

### C. Performance Parameters

Our simulation focuses on analyzing main QoS parameters for WiMAX Network, especially average throughput, packet loss rate, average delay and average jitter.

## VI. SIMULATION RESULTS AND ANALYSIS

In this paper, we perform various simulations in order to analyse and compare the proposed QoE scheduler algorithm with the NIST scheduler with respect to main QoS parameters, namely, average throughput, packet loss rate, average delay and average jitter in WiMAX network using UGS service class.

In the figure 4, we observe that the average throughput for the NIST scheduler algorithm outperform the QoE-based scheduler for all flows. Indeed, average throughput values for NIST scheduler still higher compared with the QoE-based scheduler ones.

The scheduler that takes into account the QoE varied the throughput for different users so as to match with the minimum subjective rate requirements of each user in order to reduce delays, jitter and packet loss rate.

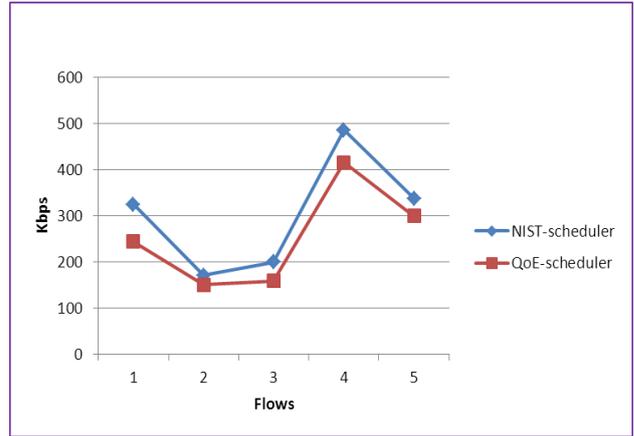

Fig 4. Average Throughput

Figure 5 shows the improvement obtained by applying the QoE-based scheduler algorithm to the packet loss rate for all flows, in general the packet loss rate is reduced. In the case of flow 5, the values are similar.

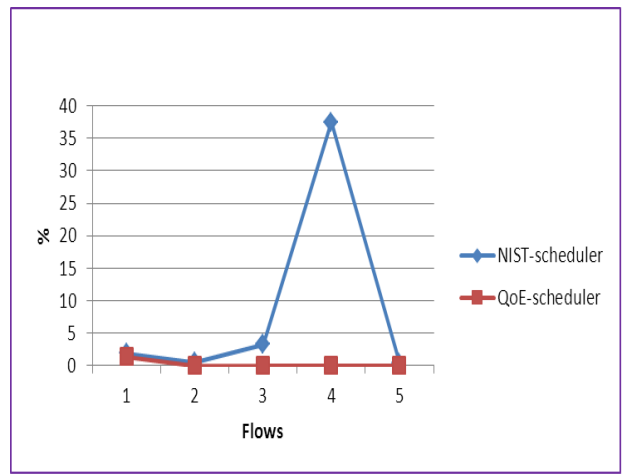

Fig 5. Packet loss rate

It can be observed from the Figure 6 that the proposed QoE-based scheduler outperforms the basic NIST scheduler in term of average jitter, values of jitter using QoE-based scheduler are extremely low compared with the NIST one. In the case of flow number five, the two schedulers have the same values of average jitter.

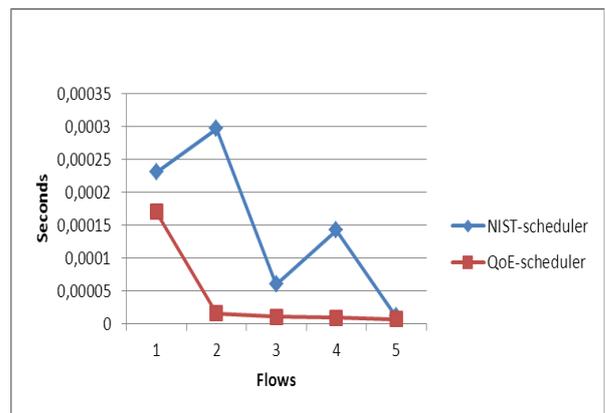

Fig 6. Average Jitter

As shown in the figure 7, the average transmission delay of packets is reduced while using the QoE-based scheduler, in the case of flow number five, the two schedulers have the same values of average delay.

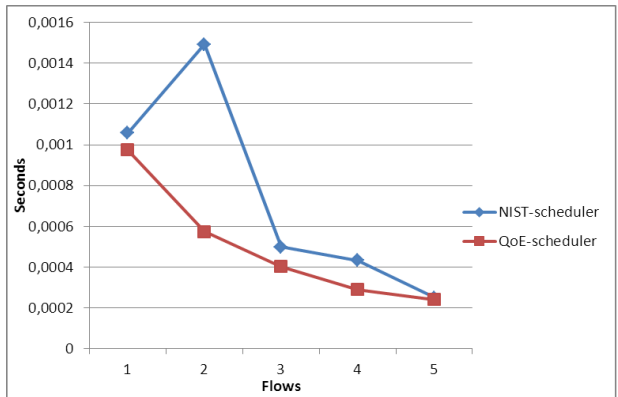

Fig 7. Average Delay

## VII. CONCLUSION

In this paper, we have used a QoE-based scheduling algorithm in which depending on whether there is a packet loss, the system reduces the transmission rate of each connection in order to match with the minimum allowed requirement of transmission rate (minimum subjective requirement of the user).

The simulations carried out show that the use of different levels of MOS enhances the QoE provided to users of WiMAX network. The proposed QoE-based scheduling algorithm significantly reduces packet loss, jitter and delay while using UGS service class.

As a future work we may extend this study by taking in consideration other service class and other subjective parameters to handle VoIP traffic.

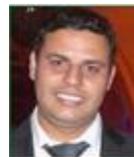

**Tarik ANOUARI** has a High Specialized Study Degree (DESS) option Information Systems Engineering, from the University of Cadi Ayyad, Faculty of Sciences Semlalia, Marrakesh, Morocco. Since November 2006 he has been working as Engineer Analyst Developer in the Deposit and Management fund (CDG), Rabat, Morocco. Currently, he is working toward his Ph.D. at the Faculty of Sciences and Techniques, Settat. His current research interests Simulation Network Performance, Network Protocols, Mobile Broadband Wireless and Analysis of Quality of experience in Next Generation Networks.

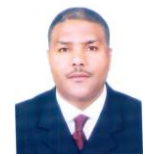

Dr. **Abdelkrim HAQIQ** has a High Study Degree (DES) and a PhD (Doctorat d'Etat), both in Applied Mathematics, option modeling and performance evaluation of computer communication networks, from the University of Mohamed V, Agdal, Faculty of Sciences, Rabat, Morocco. Since September 1995 he has been working as a Professor at the department of Mathematics and Computer at the Faculty of Sciences and Techniques, Settat, Morocco. He is the Director of Computer, Networks, Mobility and Modeling laboratory. He is also a General Secretary of e-Next Generation Networks Research Group, Moroccan section.

Dr. **Abdelkrim HAQIQ** is actually Co-Director of a NATO multi-year project entitled "Cyber Security Assurance Using Cloud-Based Security Measurement System" in collaboration with Duke University, USA, Arizona State University, USA and Canterbury University, Christchurch, New Zealand.

Dr. **Abdelkrim HAQIQ**'s interests lie in the areas of applied stochastic processes, stochastic control, queuing theory, game theory and their applications for modeling/simulation and performance analysis of computer communication networks. He is the author and co-author of more than 60 papers (international journals and conferences/workshops). He was the Chair of the second international conference on Next Generation Networks and Services, held in Marrakech, July, 8- 10, 2010 and a TPC co-chair of the fourth international conference on Next Generation Networks and Services, held in Portugal, December, 2 - 4, 2012. He was also an International Steering Committee Chair of the international conference on Engineering Education and Research 2013, iCEER2013, held in Marrakesh, July, 1st –5th, 2013. Dr. **Abdelkrim HAQIQ** is also a TPC member and a reviewer for many international conferences. He was also a Guest Editor of a special issue on Next Generation Networks and Services of the International Journal of Mobile Computing and Multimedia Communications (IJMCMC), July-September 2012, Vol. 4, No. 3. He is also a Guest Editor of a special issue of the Journal of Mobile Multimedia (JMM), Vol. 9, No.3&4, 2014.

From January 1998 to December 1998 he had a Post-Doctoral Research appointment at the department of Systems and Computers Engineering at Carleton University in Canada. He also has held visiting positions at the High National School of Telecommunications of Paris, the Universities of Dijon and Versailles St-Quentin-en-Yvelines in France, the University of Ottawa in Canada and the FUCAM in Belgium.